\documentclass[%
 reprint,
 superscriptaddress,
 amsmath,amssymb,
 prl,
floatfix,
]{revtex4-2}

\usepackage{graphicx}% Include figure files
\usepackage{dcolumn}% Align table columns on decimal point
\usepackage{bm}% bold math
\usepackage{array}% table things
\usepackage[hidelinks]{hyperref}% add hypertext capabilities
\usepackage[
separate-uncertainty = true,
multi-part-units=single
]{siunitx}
\usepackage{xcolor}
\usepackage[version=4]{mhchem}
\usepackage[normalem]{ulem}
\usepackage{tablefootnote}

\newcommand\sonec{S1${c}$}
\newcommand\stwoc{S2${c}$}

\DeclareSIUnit\keVnr{keV_{nr}}
\DeclareSIUnit\keVee{keV_{ee}}

\newcommand\tritium{tritium}
\newcommand\ArThreeSeven{\ce{^{37}Ar}}
\newcommand\KrEightThreem{\ce{^{83m}Kr}}

\newcommand\XeOneTwoSeven{\ce{^{127}Xe}}
\newcommand\XeOneTwoNine{\ce{^{129}Xe}}
\newcommand\XeOneTwoNinem{\ce{^{129m}Xe}}
\newcommand\XeOneThreeOnem{\ce{^{131m}Xe}}
\newcommand\XeOneThreeOne{\ce{^{131}Xe}}
\newcommand\supplementalurl{https://doi.org/10.1103/PhysRevLett.131.041002}

\begin{document}

\preprint{APS/123-QED}

\title{First Dark Matter Search Results from the LUX-ZEPLIN (LZ) Experiment}

% 1 
\author{J.~Aalbers}
\affiliation{SLAC National Accelerator Laboratory, Menlo Park, CA 94025-7015, USA}
\affiliation{Kavli Institute for Particle Astrophysics and Cosmology, Stanford University, Stanford, CA  94305-4085 USA}

% 2 
\author{D.S.~Akerib}
\affiliation{SLAC National Accelerator Laboratory, Menlo Park, CA 94025-7015, USA}
\affiliation{Kavli Institute for Particle Astrophysics and Cosmology, Stanford University, Stanford, CA  94305-4085 USA}

% 3 
\author{C.W.~Akerlof}
\affiliation{University of Michigan, Randall Laboratory of Physics, Ann Arbor, MI 48109-1040, USA}

% 4 
\author{A.K.~Al Musalhi}
\affiliation{University of Oxford, Department of Physics, Oxford OX1 3RH, UK}

% 5 
\author{F.~Alder}
\affiliation{University College London (UCL), Department of Physics and Astronomy, London WC1E 6BT, UK}

% 6 
\author{A.~Alqahtani}
\affiliation{Brown University, Department of Physics, Providence, RI 02912-9037, USA}

% 7 
\author{S.K.~Alsum}
\affiliation{University of Wisconsin-Madison, Department of Physics, Madison, WI 53706-1390, USA}

% 8 
\author{C.S.~Amarasinghe}
\affiliation{University of Michigan, Randall Laboratory of Physics, Ann Arbor, MI 48109-1040, USA}

% 9 
\author{A.~Ames}
\affiliation{SLAC National Accelerator Laboratory, Menlo Park, CA 94025-7015, USA}
\affiliation{Kavli Institute for Particle Astrophysics and Cosmology, Stanford University, Stanford, CA  94305-4085 USA}

% 10 
\author{T.J.~Anderson}
\affiliation{SLAC National Accelerator Laboratory, Menlo Park, CA 94025-7015, USA}
\affiliation{Kavli Institute for Particle Astrophysics and Cosmology, Stanford University, Stanford, CA  94305-4085 USA}

% 11 
% ...

\collaboration{The LUX-ZEPLIN (LZ) Collaboration}

\date{\today}% It is always \today, today,

\begin{abstract}
The LUX-ZEPLIN experiment is a dark matter detector centered on a
dual-phase xenon time projection chamber operating at the Sanford Underground Research Facility in Lead, South Dakota, USA. This Letter reports results from LUX-ZEPLIN's first search for weakly interacting massive particles (WIMPs) with an exposure of 60~live days using a fiducial mass of \SI{5.5}{t}. A profile-likelihood ratio analysis shows the data to be consistent with a background-only hypothesis, setting new limits on spin-independent WIMP-nucleon, spin-dependent WIMP-neutron, and spin-dependent WIMP-proton cross sections for WIMP masses above \SI{9}{GeV/c\squared}. The most stringent limit is set for spin-independent scattering at \SI{36}{GeV/c\squared}, rejecting cross sections above \SI{9.2e-48}{cm\squared} at the \SI{90}{\percent} confidence level. 

% Abstract below is for arxiv metadata submission
%The LUX-ZEPLIN (LZ) experiment is a dark matter detector centered on a dual-phase xenon time projection chamber operating at the Sanford Underground Research Facility in Lead, South Dakota, USA. This Letter reports results from LZ's first search for Weakly Interacting Massive Particles (WIMPs) with an exposure of 60 live days using a fiducial mass of 5.5 t. A profile-likelihood ratio analysis shows the data to be consistent with a background-only hypothesis, setting new limits on spin-independent WIMP-nucleon, spin-dependent WIMP-neutron, and spin-dependent WIMP-proton cross-sections for WIMP masses above 9 GeV/c$^2$. The most stringent limit is set at 30 GeV/c$^2$, excluding cross sections above 6.5$\times 10^{-48}$ cm$^2$ at the 90\% confidence level. 
\end{abstract}

\keywords{Dark Matter, Direct Detection, Xenon}%Use showkeys class option if keyword
\maketitle

%\tableofcontents

There is abundant astrophysical evidence for the existence of dark matter~\cite{Aghanim2020, SofueRubin, harvey2015nongravitational, ARBEY2021103865}, a nonrelativistic and nonbaryonic matter component of the Universe that has so far eluded direct detection through interaction with ordinary matter~\cite{schumann2019direct}. Weakly interacting massive particles (WIMPs), which obtain their relic abundance by thermal freeze-out through weak interactions~\cite{Lee1977ua}, are postulated in a wide variety of viable extensions to the standard model of particle physics~\cite{RevModPhys.90.045002, Billard2022,akerib2022snowmass2021}. They are a leading candidate to explain dark matter, despite strong constraints from many searches completed and ongoing at colliders~\cite{aaboud2018search, sirunyan2018search, aaboud2018searchz, sirunyan2018searchz, universe4110131}, with telescopes~\cite{abe2020indirect,choi2015search, abbasi2022search, albert2017results,  cuoco2017novel, cui2017possible, Albert_2017}, and in underground laboratories~\cite{abdelhameed2019first, agnes2018low,amole2019dark,collaboration2018dark, PhysRevLett.127.261802,deapcollaborationSearchDarkMatter2019, akerib2017results, Agnese:2015ywx}. 
This Letter reports the first search for dark matter from the LUX-ZEPLIN (LZ) experiment, with the largest target mass of any WIMP detection experiment to date.

The LZ experiment~\cite{LZExperiment, lztdr} is located \SI{4850}{ft} underground in the Davis Cavern at the Sanford Underground Research Facility (SURF) in Lead, South Dakota, USA, shielded by an overburden of \SI{4300}{m} water-equivalent~\cite{Heise:2021eym}. It is a low-background, multidetector experiment centered on a dual-phase time projection chamber (TPC) mounted in a double-walled titanium cryostat~\cite{AKERIB20171} filled with \SI{10}{t} of liquid xenon (LXe). The TPC is a vertical cylinder approximately \SI{1.5}{m} in diameter and height, lined with reflective polytetrafluoroethylene, and instrumented with 494 3{\nobreakdash-}inch photomultiplier tubes (PMTs) in two arrays at top and bottom. Energy depositions above approximately \SI{1}{keV} in the \SI{7}{t} active xenon region produce two observable signals: vacuum ultraviolet (VUV) scintillation photons (S1) and ionization electrons that drift under a uniform electric field to the liquid surface, where they are extracted and produce secondary scintillation in the xenon gas (S2). The ratio of S2 to S1 differentiates interactions with a xenon nucleus (producing a nuclear recoil, or NR) from interactions with the atomic electron cloud (producing an electron recoil, or ER). 
 
The TPC is surrounded by two detectors, which provide veto signals to reject internal and external backgrounds. A LXe ``skin'' detector between the TPC field cage and the cryostat wall is instrumented with 93 1{\nobreakdash-}inch and 38 2{\nobreakdash-}inch PMTs.  The outer detector (OD) is a near-hermetic system of acrylic tanks containing \SI{17}{t} of gadolinium-loaded (\SI{0.1}{\percent} by mass) liquid scintillator~\cite{Haselschwardt:2018vmp} to detect neutrons. The entire LZ detector system is in a tank filled with \SI{238}{t} of ultrapure water to shield from the ambient radioactive background, and 120 8-inch PMTs are submersed in the water to record OD and water Cherenkov signals.

The data reported here were collected from December 23, 2021 to May 11, 2022, under stable detector conditions. The cathode and gate electrodes \cite{LINEHANgrids} established a drift field of \SI{193}{V/cm}, determined by electrostatic simulation to vary by \SI{4}{\percent} over the volume considered in this analysis. The gate and anode electrodes established a gas extraction field of \SI{7.3}{kV/cm} at radial position $r=0$. Twelve TPC and two skin PMTs, with no specific position correlation, developed malfunctioning connections or excessive noise during commissioning and were disabled prior to the run. The temperature and pressure of the LXe were stable to within 0.2$\%$, at \SI{174.1}{\kelvin} (at the TPC bottom) and \SI{1.791}{bar(a)}. The liquid level was stable to within \SI{10}{\micro\meter}, measured by precision capacitance sensors. The full xenon complement of \SI{10}{t} was continuously purified at \SI{3.3}{t/day} through a hot getter system, and the observed electron lifetime against attachment on electronegative impurities was 
between \SI{5000}{\micro\s} and \SI{8000}{\micro\s}, much longer than the \SI{951}{\micro\s} maximum drift time in the TPC. 

The data acquisition (DAQ) system records events triggered by a digital filter sensitive to S2 signals in the TPC, reaching full efficiency for S2 pulses with six extracted electrons at a typical rate of \SI{5}{Hz}. A time window of \SI{2}{ms} before and \SI{2.5}{ms} after each trigger is recorded, constituting an event. Zero-suppressed waveforms from all PMT channels, including low- and high-gain amplification paths for TPC and OD PMTs, are recorded for every trigger with single photoelectron efficiencies averaging \SI{94}{\percent}, \SI{86}{\percent}, and $>$\SI{95}{\percent} for the TPC, skin, and OD PMTs, respectively.

Event properties are reconstructed through analysis of the PMT waveform shapes, timings, and distributions. Raw waveform amplitudes are normalized by the PMT and amplifier gains and summed separately within the TPC, skin, and OD. Integrated waveform area is reported in photons detected (phd) at each PMT, accounting for the double photoelectron effect in response to vacuum ultraviolet photons~\cite{faham2015measurements, LOPEZPAREDESdpe}. Pulse boundaries are identified on the summed waveforms using filters tuned for prototypical pulse shapes in each detector. Pulses in the TPC are further classified as S1 or S2 based on their hit pattern and pulse shape. S1 pulses are required to have signals above the electronic noise threshold in at least three PMTs. The time ordering of the most prominent S1 and S2 pulses in each event is then used to identify single-scatter (one S1 preceding one S2) and multiscatter (one S1 preceding multiple S2s) events. The transverse $(x,y)$ location of events is determined by the PMT hit pattern of S2 light from the extracted electrons, using the \textsc{Mercury} algorithm~\cite{Mercury}. The algorithm was tuned using uniformly distributed radioactive sources in the TPC and has a 1$\sigma$ resolution of \SI{4}{mm} for S2 signals of \SI{3000}{phd}. The resolution worsens by approximately a factor of 2~near the TPC wall due to asymmetric light collection at the TPC edge. The location along the cylinder ($z$) axis is inferred from the drift time, and has a 1$\sigma$ resolution of \SI{0.7}{mm} for events near the cathode electrode.

LZ uses radioactive sources to correct for spatial variation in response across the TPC and to calibrate the detector response to ER and NR events. ER calibration events are obtained using dispersed sources \KrEightThreem\ and \XeOneThreeOnem\ before and during the WIMP search and tritiated methane (CH$_3$T) postsearch. The tritium source is important for understanding the response to low-energy ER events, the most prominent background component in the run. Localized NR calibration events are created using a deuterium-deuterium (DD) generator that emits monoenergetic \SI{2.45}{MeV} neutrons~\cite{adelphi, Verbus_others_2017, LUX_DD} along a conduit through the water tank at approximately 10 cm below the liquid surface and AmLi sources~\cite{MOZHAYEVAmLi} deployed between the walls of the cryostat vessels in three azimuthal positions and three $z$ positions, a total of nine positions.

Using the dispersed sources, the S1 signal is normalized to the geometric center of the detector, using a correction in $x$, $y$, and drift time; this normalized value is called \sonec. The S2 signal is normalized to a signal at the radial center and top (shortest drift time) of the detector; this normalized value is called \stwoc. The size of the S1 corrections is on average \SI{9}{\percent} and comes primarily from variations in light collection efficiency and PMT quantum efficiency. The size of the S2 corrections is on average \SI{11}{\percent} in the $(x, y)$ plane and comes primarily from nonoperational PMTs and extraction-field nonuniformity caused by electrostatic deflection of the gate and anode electrodes. The S2 correction in $z$ is due to electron attachment on impurities and averages \SI{7}{\percent}.  Corrected parameters are uniform across the TPC to within $3$\%. 

\begin{figure}[htbp]
  \centering
  \includegraphics[ width=\linewidth]{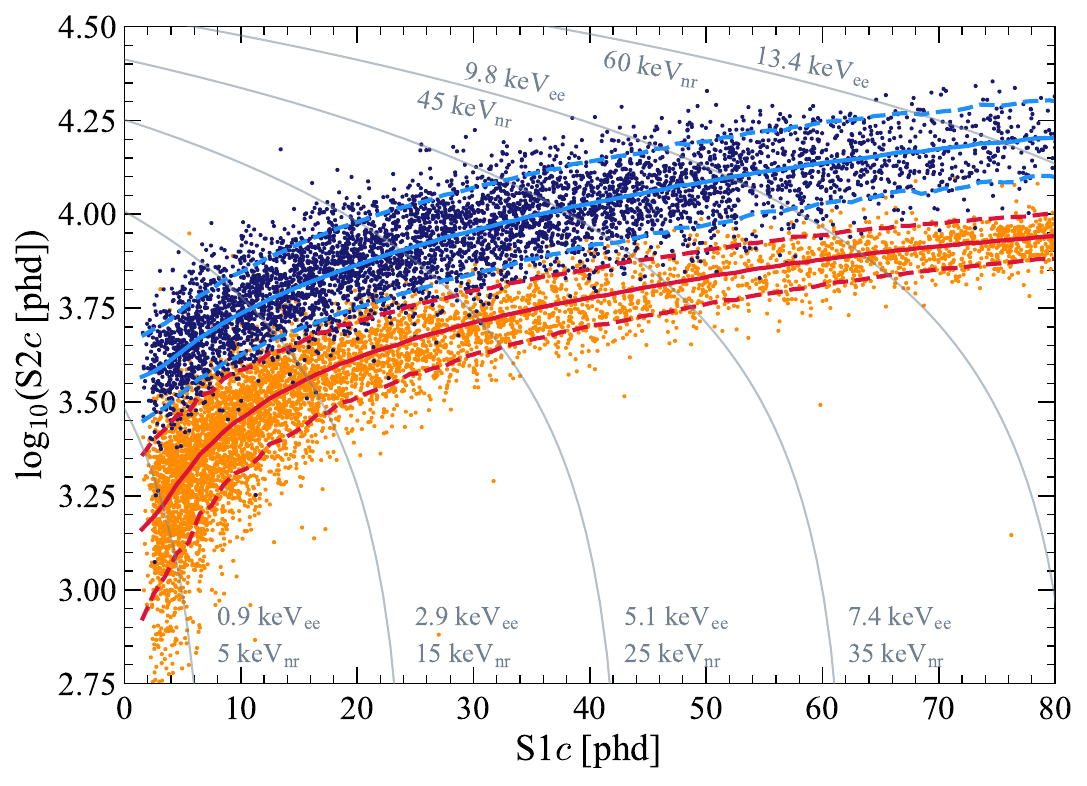}
  \caption{Calibration events in $\log_{10}$\stwoc-\sonec\ for the tritium\ source (dark blue points, 5343 events) and the DD neutron source (orange points, 6324 events). Solid blue (red) lines indicate the median of the ER (NR) simulated distributions, and the dotted lines indicate the {\SI{10}{\percent}} and {\SI{90}{\percent}} quantiles. Thin gray lines show contours of constant electron-equivalent energy (keV$_{\text{ee}}$) and nuclear recoil energy (keV$_{\text{nr}}$).}
  \label{fig:S1S2}
\end{figure}

To reproduce the TPC response to ER and NR events, detector and xenon response parameters of the \textsc{nest} 2.3.7~\cite{szydagis_m_2022_6534007} ER model are tuned to match the median and widths of the \tritium\ calibration data in log$_{10}$\stwoc-\sonec\, space, and to match the reconstructed energies of the \KrEightThreem~(\SI{41.5}{keV}), \XeOneTwoNinem~(\SI{236}{keV}), and \XeOneThreeOnem~(\SI{164}{keV}) peaks. The photon detection efficiency $g_1$ is determined to be \SI{0.114(2)}{phd/photon} and the gain of the ionization channel $g_2$ to be \SI{47.1(11)}{phd/electron}~\footnote{See Supplemental Material at \url{\supplementalurl} for the full description of the \textsc{nest} parameters and a header file to configure the software for this model}. The \tritium\ data are best modeled with the \textsc{nest} recombination skewness model~\cite{akerib2020discrimination} disabled, and comparisons between the tuned model and tritium data using several statistical tests show consistency throughout the full tritium ER distribution~\cite{kolmogorov1933sulla,smirnov1939estimation, andersondarling, shapiro1965analysis}. The \textsc{nest}  ER model also includes effects from electron capture decays~\cite{temples2021measurement} when making predictions from electron capture background sources. The parameters of the ER model were propagated to the \textsc{nest} NR model and found to be in good agreement with  DD calibration data, matching NR band means and widths to better than \SI{1}{\percent} and \SI{4}{\percent} in log$_{10}$\stwoc, respectively. Further checks comparing DD and AmLi neutron calibrations agree to 1\%. Figure \ref{fig:S1S2} shows the \tritium\ and DD neutron data compared to the calibrated model. 

The WIMP signal considered in this analysis is expected to produce low-energy, single-scatter NR signals uniformly distributed in the TPC, with no additional signals in the TPC, skin, or OD. 
The following strategy is used to obtain a clean sample of such events: exclude time periods of elevated TPC activity or electronics interference; remove multiscatter interactions in the TPC; remove events outside an energy region of interest (ROI); remove events due to accidental coincidence of S1 and S2 pulses; remove events with coincident signals in the TPC and skin or OD; remove events near the TPC active volume boundaries.
Methods of bias mitigation that involve obscuring the data, such as blinding the signal region or adding fake events (``salting''), were avoided to allow control over larger sources of systematic errors that may be presented by a new detector. To mitigate bias in this result, all analysis cuts were developed and optimized on sideband selections and calibration data.

The search dataset totals 89~live days after removing periods for detector maintenance and calibration activity, as well as a \SI{3}{\percent} loss due to DAQ dead time and a \SI{7}{\percent} loss to periods excised due to anomalous trigger rates. Because dual-phase xenon TPCs experience elevated rates of activity after large S2 pulses~\cite{akerib2017results, collaboration2018dark,sorensen2017electron, LUXelectronbkgs}, a time hold-off is imposed to remove data taken after large S2s and after cosmic-ray muons traversing the TPC. These omissions result in a final search live time of \SI{60(1)}{d} where a WIMP interaction could be reconstructed. 
In future searches, the hold-off can be relaxed by optimization with respect to analysis cuts and detector operating conditions.

The ROI is defined as \sonec~in the range $3-80$\,phd, uncorrected S2 greater than \SI{600}{phd} ($>$10 extracted electrons), and \stwoc~less than $10^5$\,phd, ensuring that signal efficiencies are well understood and background ER sources are well calibrated by the tritium data.
Events classified as multiple scatters in the TPC are removed, as are events with poor reconstruction due to noise, spurious pulses, or other data anomalies.

A suite of analysis cuts targets accidental coincidence events, henceforth called ``accidentals,'', where an isolated S1 and an isolated S2 are accidentally paired to mimic a physical single-scatter event. Isolated S1s can be generated from sources such as particle interactions in charge-insensitive regions of the TPC, Cherenkov and fluorescent light in detector materials, or dark-noise pileup. Isolated S2s can be generated from sources such as radioactivity or electron emission from the cathode or gate electrodes, particle interactions in the gas phase or in the liquid above the gate electrode, or drifting electrons trapped on impurities and released with $\mathcal{O}$(\SI{100}{ms}) time delay \cite{LUXelectronbkgs}.
Analysis cuts to remove accidentals target individual sources of isolated S1s and S2s using the expected behavior of the S1 and S2 pulses with respect to quantities such as drift time, top-bottom asymmetry of light, pulse width, timing of PMT hits within the pulse, and hit pattern of the photons in the PMT arrays. The cuts remove $>$\SI{99.5}{\percent} of accidentals, measured using single-scatter-like events with unphysical ($>$\SI{951}{\micro\second}) drift time and events generated by random matching of isolated S1 and S2 populations. 

\begin{figure}[htbp]
  \centering
  \includegraphics[ width=\linewidth]{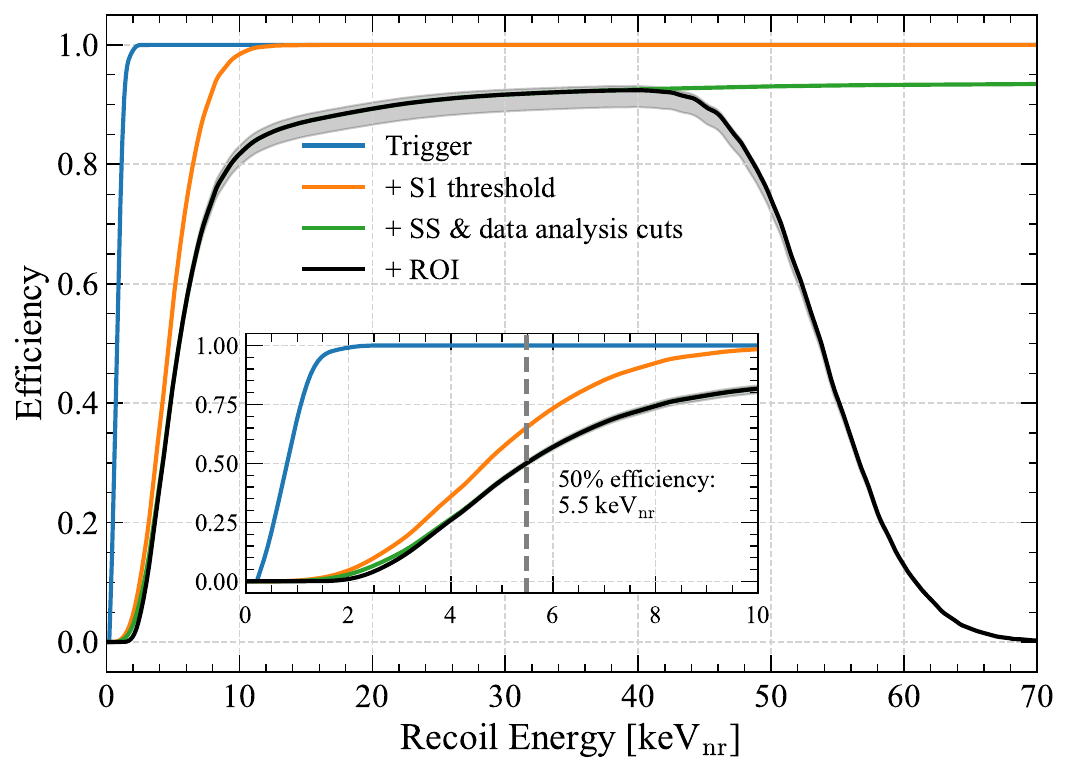}
  \caption{Signal efficiency as a function of NR energy for the trigger (blue), the threefold coincidence and $>$\SI{3}{phd} threshold on S1c (orange), single-scatter (SS) reconstruction and analysis cuts (green), and the search ROI in S1 and S2 (black). The inset shows the low-energy behavior, with the dotted line at \SI{5.5}{keV_{\text{nr}}} marking  \SI{50}{\percent} efficiency. 
  The error band (gray) is assessed using AmLi and \tritium\ data as discussed in the text.
  }
  \label{fig:NRacc}
\end{figure}

Data-driven signal efficiencies for the trigger, reconstruction, and analysis cuts are shown in Fig.~\ref{fig:NRacc}. The DAQ trigger efficiency is determined from DD data by comparing the external trigger of the generator against the TPC S2 trigger logic, and is confirmed using randomly triggered events collected throughout the search.
The reconstruction efficiency for low-energy NR events is evaluated by comparing the reconstruction results against a large set of events manually identified as single scatter in DD data. An additional reconstruction inefficiency due to S2 splitting for long drift times for low numbers of extracted electrons is accounted for with simulation.
Analysis cut efficiency is not determined directly from neutron calibration data as they do not cover the spatial extent of the TPC and are contaminated by a high rate of single photons and electrons. 
Instead, the efficiency throughout the full analysis volume is evaluated using \tritium\ data for analysis cuts targeting S1 pulses and the combination of \tritium\ and AmLi data for those targeting S2 pulses. Composite NR-like waveforms are generated using \tritium\ single scatters with their S2 pulses replaced by smaller pulses from other \tritium\ or AmLi events (an ``AmLi-\tritium'' dataset).  The uncertainty on the NR signal efficiency is the larger of the $\pm1\sigma$ statistical fluctuation of the AmLi-\tritium\ dataset and the difference between the AmLi-\tritium\ dataset and a pure AmLi dataset. The uncertainty is \SI{3}{\percent} for nuclear recoil energies $>$\SI{3.5}{keV_{\text{nr}}}, increasing to \SI{15}{\percent} at \SI{1}{keV_{\text{nr}}}.

Events with coincident activity in the TPC and skin or OD are removed to reduce backgrounds producing $\gamma$ rays and neutrons. To mitigate backgrounds associated with $\gamma$ rays, events with a prompt signal in the OD (skin) within $\pm$\SI{0.3}{\micro\s} ($\pm$\SI{0.5}{\micro\s}) of the TPC S1 pulse are removed. Neutrons can thermalize in detector materials and those that capture on  hydrogen or gadolinium in the OD can be tagged by an OD pulse of greater than $\sim$\SI{200}{keV} within \SI{1200}{\micro\s} after the TPC S1. A selection on large skin pulses in the same time window additionally tags $\gamma$ rays returning to the xenon from an OD capture process. AmLi calibration sources placed at the nine locations close to the TPC are used to determine a position-averaged neutron tagging efficiency of \SI{89(3)}{\percent} for TPC single-scatters in the nuclear recoil band. Background data is used to determine a false veto rate of $5\%$ due to accidental activity in the OD during the coincidence window. Background neutrons may have a higher tagging efficiency due to their harder energy spectrum and coincident $\gamma$-ray emission.

\begin{figure}[tbp]
 \centering
 \includegraphics[ width=\linewidth]{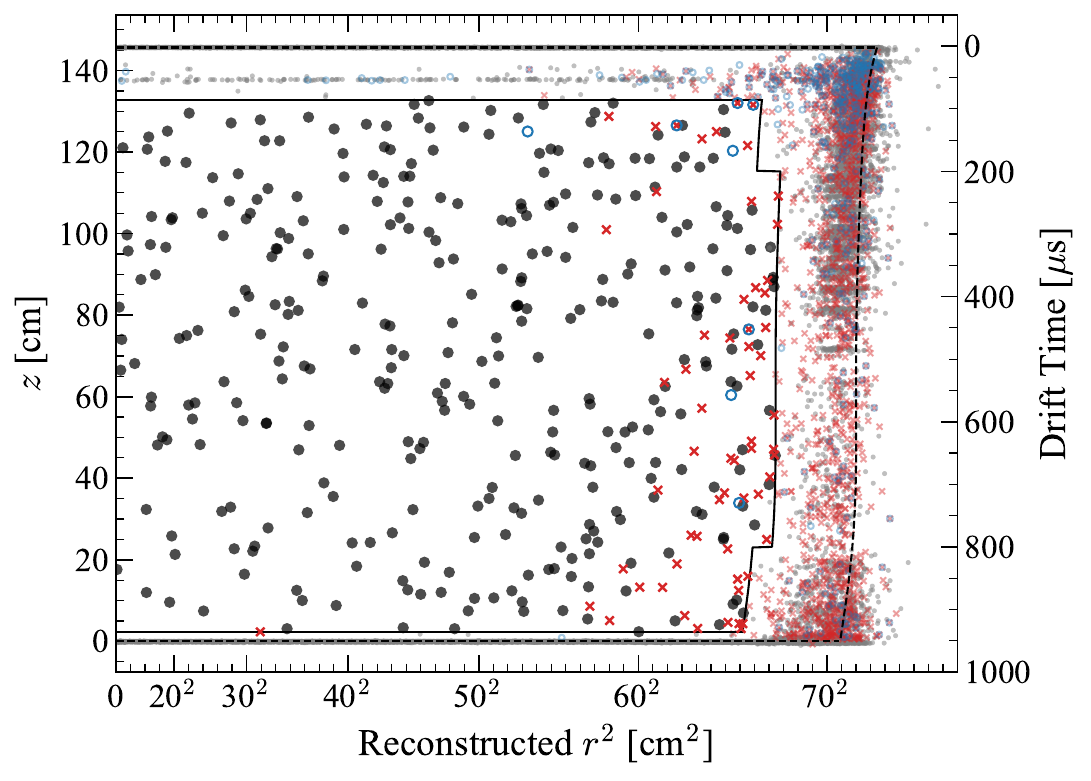}
 \caption{Data in reconstructed $r^2$ and $z$ after all analysis cuts. Black (gray) points show the data inside (outside) the FV. Red crosses and blue circles show events vetoed by a prompt LXe skin or OD signal, respectively. The solid line shows the FV definition, and the dashed line shows the extent of the active TPC. Field nonuniformities cause the reconstructed $r$ position of the active volume boundary to vary as a function of z. Events with drift time of approximately \SI{50}{\micro\s} are from recoils in the gas that produce S1 and S2 pulses with a fixed time separation.}
 \label{fig:r2z}
\end{figure}

Finally, events outside a central fiducial volume (FV) are removed to reject external and other backgrounds that concentrate near the TPC boundaries, as shown in Fig.~\ref{fig:r2z}. Events at high radius have reduced position reconstruction resolution due to reduced S2 light collection efficiency and charge-loss effects within a few millimeters of the polytetrafluoroethylene wall.
The radial extent of the FV and the S2 threshold are chosen simultaneously using data outside the \sonec\ ROI to eliminate events leaking into the FV due to poor position reconstruction resolution. Radially, the FV terminates at \SI{4.0}{cm} in reconstructed position from the TPC wall, with small additional volumes removed in the top (\SI{5.2}{cm} for drift time $<$\SI{200}{\micro\s}) and bottom (\SI{5.0}{cm} for drift time $>$\SI{800}{\micro\s}) corners to account for increased rates of background in those locations. Events within \SI{6.0}{cm} of the $(x, y)$ positions of two ladders of TPC field-cage resistors embedded in the TPC wall are also removed.  Vertically, events with drift times $<$\SI{86}{\micro\s} and $>$\SI{936.5}{\micro\s} are rejected, corresponding to \SI{12.8}{cm} and \SI{2.2}{cm} from the gate and cathode electrodes, respectively.
The number of remaining events from the wall entering the FV is estimated to be $<0.01$.
The xenon mass in the FV is estimated to be \SI{5.5(2)}{\tonne} using tritium data and confirmed by geometric calculation. 

\begin{figure}[tbp]
  \centering
  \includegraphics[ width=\linewidth]{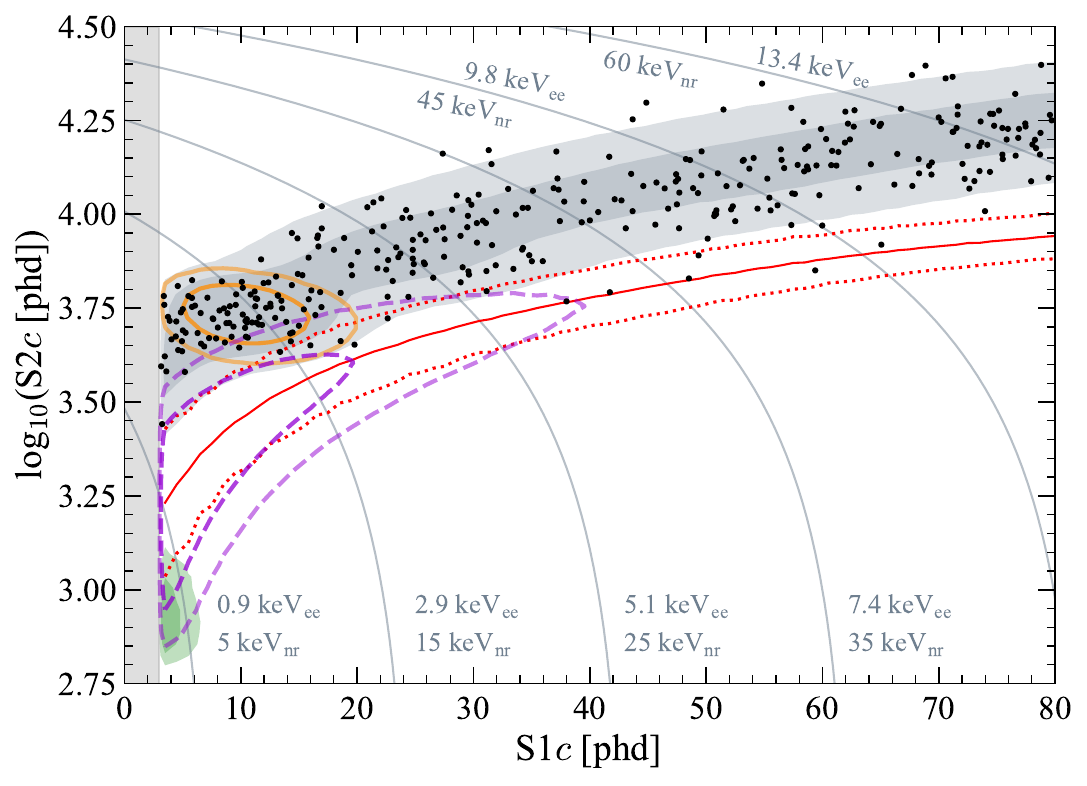}
  \caption{WIMP-search data (black points) after all cuts in $\log_{10}$\stwoc-\sonec~space. Contours enclose $1\sigma$ and $2\sigma$ of the following models: the best-fit background model (shaded gray regions), the \ArThreeSeven\ component (orange ellipses), a \SI{30}{GeV/c\squared} WIMP (purple dashed lines), and $^8$B solar neutrinos (shaded green regions).
  The red solid line indicates the NR median, and the red dotted lines indicate the \SI{10}{\percent} and \SI{90}{\percent} quantiles.
  Model contours incorporate all efficiencies used in the analysis. 
  Thin gray lines indicate contours of constant energy.
  }
  \label{fig:s1s2wimp}
\end{figure}

Figure~\ref{fig:s1s2wimp} shows the distribution in $\log_{10}$\stwoc{\nobreakdash-}\sonec\ of the 335 events~\footnote{See Supplemental Material %at \url{\supplementalurl} 
for a detailed table of events by selection.}  
passing all selections, along with contours representing a \SI{30}{GeV/c\squared} WIMP, a flat NR distribution, and the background model.
The signal model assumes spin-independent scattering from WIMPs with an isotropic Maxwell–Boltzmann velocity distribution, parameterized as in Ref.~\cite{baxter2021recommended}, with inputs from Refs.~\cite{LEWIN199687,smith2007rave,mccabe2014earth, schonrich2010local,blandhawthorn2016galaxy, gravity2021improved}.
The WIMP model has an approximately exponentially decreasing energy spectrum with shape that depends on the mass of the WIMP~\cite{LEWIN199687}.

The background model in this analysis consists of nine components, grouped according to their spectra in the ROI or the uncertainty on their rate. Table~\ref{tab:bkgds} lists the expected number of events from each component.

\begin{table}[htbp]
    \caption{Number of events from various sources in the \SI{60}{d}$\times$\SI{5.5}{t} exposure~\cite{aalbers2022background}. The middle column shows the predicted number of events with uncertainties as described in the text. The uncertainties are used as constraint terms in a combined fit of the background model plus a \SI{30}{GeV/c\squared} WIMP signal to the selected data, the result of which is shown in the right column. $^{37}$Ar and detector neutrons have non-Gaussian prior constraints and are totaled separately. Values at zero have no lower uncertainty due to the physical boundary.}
    \label{tab:bkgds}
    \centering
    \begin{tabular}{>{\centering}p{3.3cm}>{\centering}p{2.38cm}>{\centering}p{2.3cm}}
    \tabularnewline
    \hline
    \hline
    Source & Expected Events & Fit Result\tabularnewline
    \hline
    $\beta$ decays + Det. ER &\centering 215 $\pm$ 36\phantom{0} &\centering 222 $\pm$ 16\phantom{0}  \tabularnewline
    $\nu$ ER& 27.1 $\pm$ 1.6\phantom{0} & 27.2 $\pm$ 1.6\phantom{0} \tabularnewline
    \XeOneTwoSeven & 9.2 $\pm$ 0.8 & 9.3 $\pm$ 0.8 \tabularnewline 
    \ce{^{124}Xe}& 5.0 $\pm$ 1.4 & 5.2 $\pm$ 1.4 \tabularnewline
    \ce{^{136}Xe} &	15.1 $\pm$ 2.4\phantom{0} & 15.2 $\pm$ 2.4\phantom{0} \tabularnewline
    $^8$B CE$\nu$NS & 0.14 $\pm$ 0.01 & 0.15 $\pm$ 0.01 \tabularnewline
    Accidentals	& 1.2 $\pm$ 0.3 & 1.2 $\pm$ 0.3 \tabularnewline
    \hline
    \rule{0pt}{1.0ex} Subtotal & 273 $\pm$ 36\phantom{0} 
    & 280 $\pm$ 16\phantom{0} \tabularnewline
    \hline \tabularnewline[-2.2ex]
    \ce{^{37}Ar} & [0, 288] & $52.5^{+9.6}_{-8.9}$\phantom{0} \tabularnewline[0.25ex]
    Detector neutrons &	$0.0^{+0.2 }$ & $0.0^{+0.2}$ \tabularnewline[0.25ex]
    \SI{30}{GeV/c\squared} WIMP &\centering -- &\centering $0.0^{+0.6}$ \tabularnewline[0.25ex]
    \hline
    Total & -- & 333 $\pm$ 17\phantom{0} \tabularnewline
    \hline
    \hline
    \end{tabular}\\
\end{table}

The dominant ER signal in the search comes from radioactive decay of impurities dispersed in the xenon. $^{214}$Pb from the $^{222}$Rn decay chain, $^{212}$Pb from $^{220}$Rn, and $^{85}$Kr have broad energy spectra that are nearly flat in energy across the ROI and are summed into an overall $\beta$ background. The concentrations of $^{214}$Pb (\SI{3.26}{\micro Bq/kg}) and $^{212}$Pb (\SI{0.14}{\micro Bq/kg}) are determined by fitting to energy peaks outside the ROI. 
The xenon was purified of krypton above ground using gas chromatography~\cite{akerib2018chromatography}, and an \textit{in situ} mass spectroscopy measurement of \SI{144(22)}{ppq} $^{\mathrm{nat}}$Kr~(g/g) informs the $^{85}$Kr rate estimate. 
The $\beta$ component is further combined with a small ($<$\SI{1}{\percent}) and similarly flat ER contribution from $\gamma$ rays originating in the detector components~\cite{akerib2020cleanliness} and cavern walls~\cite{AKERIB2020102391}. Solar neutrinos are also predicted to contribute a nearly flat ER spectrum in the ROI, with a rate calculated using Refs.~\cite{baxter2021recommended, Agostini2019Simultaneous, vinyoles2017new, aharmin2013combined}. As the prediction is very precise, neutrinos are kept separate from the detector $\beta$ background in this model. The naturally occurring isotopes of $^{124}$Xe (double electron capture) and $^{136}$Xe (double $\beta$ decay) contribute ER events, and the predictions are driven by the known isotopic abundances, lifetimes, and decay schemes~\cite{BerglundWieserIso,aprile2019observation,PhysRevC.89.015502}. 

Cosmogenic activation of the xenon prior to underground deployment produces short-lived isotopes that decayed during this first run, notably $^{127}$Xe (\SI{36.3}{d}) and $^{37}$Ar (\SI{35.0}{d})~\cite{TabRad_v7,TabRad_v8, PhysRevD.96.112011}. Atomic de-excitations following $^{127}$Xe L- or M-shell electron captures fall within the ROI if the ensuing $^{127}$I nuclear de-excitation $\gamma$ ray(s) escapes the TPC.  The rate of $^{127}$Xe electron captures is constrained by the rate of K-shell atomic de-excitations, which are outside the ROI.  The skin is effective at tagging the $^{127}$I nuclear de-excitation $\gamma$ ray(s), reducing this background by a factor of 5. 
The number of $^{37}$Ar events is estimated by calculating the exposure of the xenon to cosmic rays before it was brought underground, then correcting for the decay time before the search~\cite{aalbers2022cosmogenic}. A flat constraint of 0 to three times the estimate of 96 events is imposed because of large uncertainties on the prediction.

The NR background has contributions from radiogenic neutrons and coherent elastic neutrino-nucleus scattering (CE$\nu$NS) from $^8$B solar neutrinos. The prediction for the CE$\nu$NS rate, calculated as in Refs.~\cite{baxter2021recommended, Agostini2019Simultaneous, vinyoles2017new, aharmin2013combined}, is small due to the S2$>$600 phd requirement. The rate of radiogenic neutrons in the ROI is constrained using the distribution of single scatters in the FV tagged by the OD and then applying the measured neutron tagging efficiency from the AmLi calibration sources (\SI{89(3)}{\percent}). A likelihood fit of the NR component in the OD-tagged data is consistent with observing zero events, leading to a data-driven constraint of $0.0^{+0.2}$ applied to the search. This rate agrees with simulations based on detector material radioassay~\cite{akerib2020cleanliness}.

Finally, the expected distribution of accidentals is determined by generating composite single-scatter event waveforms from isolated S1 and S2 pulses and applying the WIMP analysis selections. The selection efficiency is then applied to unphysical drift time single-scatter-like events to constrain the accidentals rate.

\begin{figure}[htbp]
  \centering
  \includegraphics[ width=\linewidth]{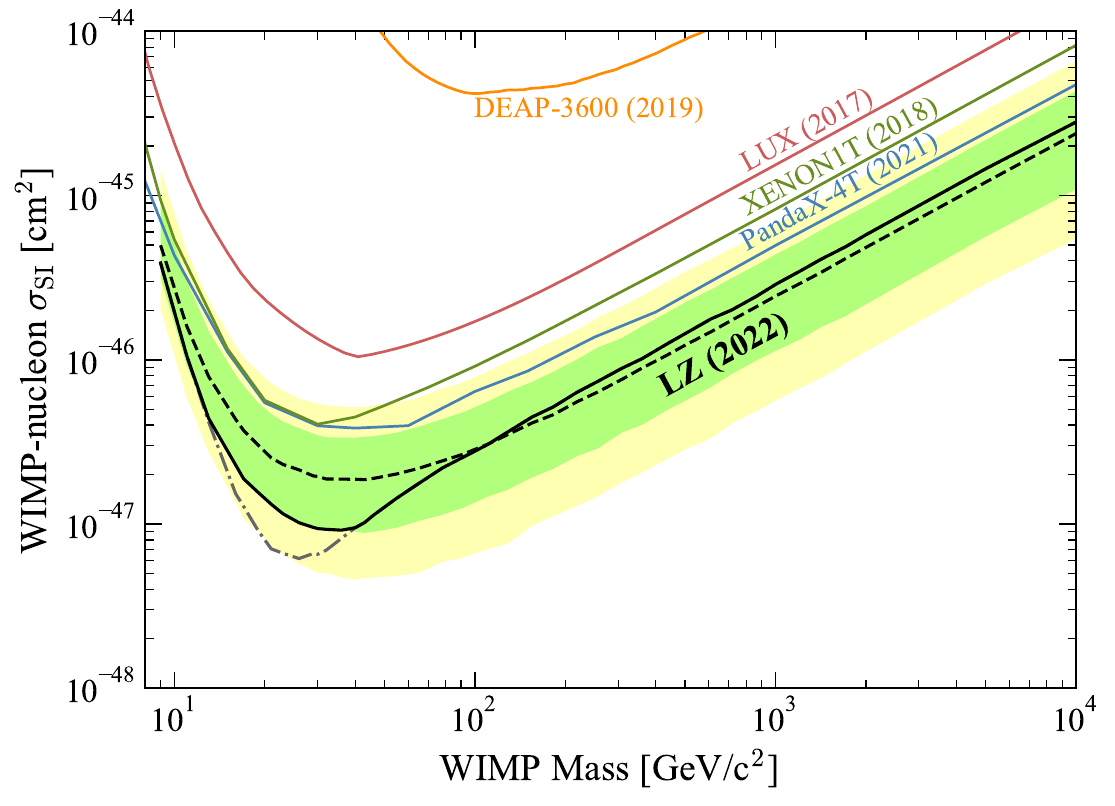}
  \caption{The \SI{90}{\percent} confidence limit (black line) for the spin-independent WIMP cross section vs.~WIMP mass. The gray dot-dash line shows the limit before applying the power constraint described in the text. The green and yellow bands are the $1\sigma$ and $2\sigma$ sensitivity bands. The dotted line shows the median of the sensitivity projection. Also shown are the PandaX-4T~\cite{PhysRevLett.127.261802}, XENON1T~\cite{collaboration2018dark}, LUX~\cite{akerib2017results}, and DEAP-3600~\cite{DEAP:2019yzn} limits. }
  \label{fig:limitplot}
\end{figure}

\begin{figure}[htbp]
  \centering
  \includegraphics[ width=\linewidth]{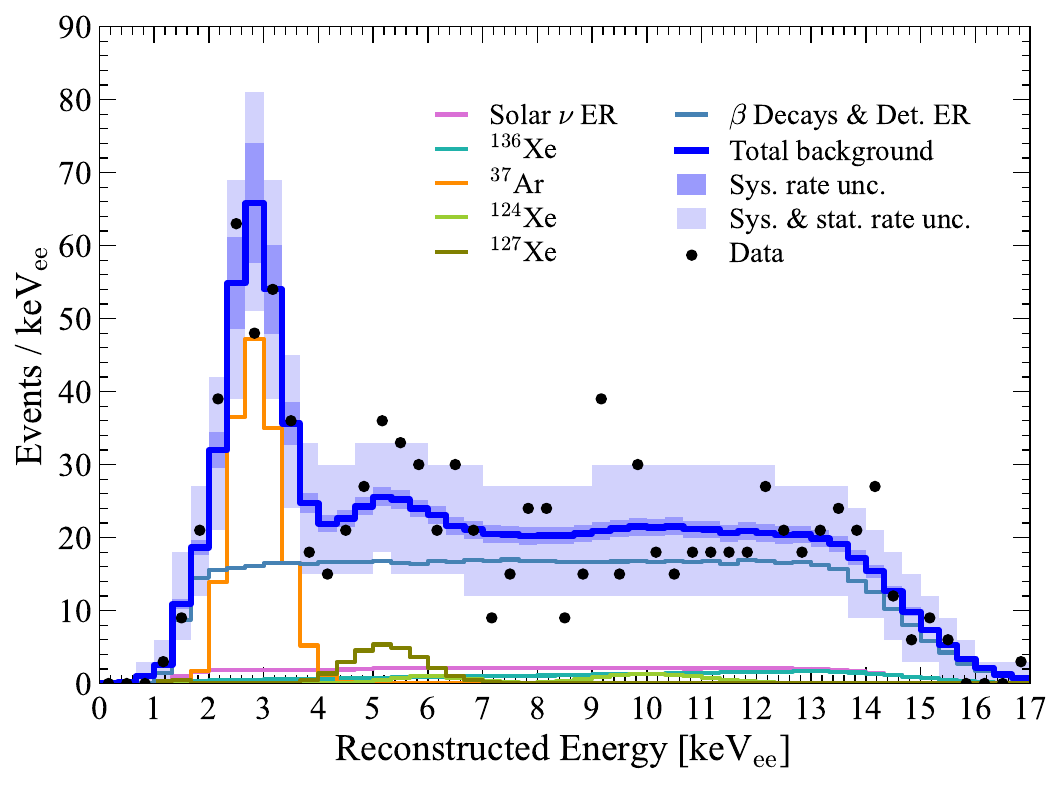}
  \caption{Reconstructed energy spectrum of the best-fit model. Data points are shown in black. The blue line shows total summed background. The darker blue band shows the model uncertainty and the lighter blue band the combined model and statistical uncertainty. Background components are shown in colors as given in the legend. Background components from $^8$B solar neutrinos and accidentals are included in the fit but are too small to be visible in the plot.}
  \label{fig:energy}
\end{figure}

Statistical inference of WIMP scattering cross section and mass is performed with an extended unbinned profile likelihood statistic in the $\log_{10}$\stwoc-\sonec\ observable space, with a two-sided construction of the \SI{90}{\percent} confidence bounds~\cite{baxter2021recommended}.
Background and signal component shapes are modeled in the observable space using the \textsc{geant4}-based package \textsc{baccarat}~\cite{akerib2021simulations,ALLISON2016186} and a custom simulation of the LZ detector response using the tuned \textsc{nest} model. The background component uncertainties are included as constraint terms in a combined fit of the background model to the data, the result of which is also shown in Table~\ref{tab:bkgds}. 

Above the smallest tested WIMP mass of \SI{9}{GeV/c\squared}, the best-fit number of WIMP events is zero, and the data are thus consistent with the background-only hypothesis.  Figure~\ref{fig:limitplot} shows the \SI{90}{\percent} confidence level upper limit on the spin-independent WIMP-nucleon cross section $\sigma_{\mathrm{SI}}$ as a function of mass. For WIMP masses between \SI{13}{GeV/c\squared} and \SI{36}{GeV/c\squared}, background fluctuations produce a limit that is substantially smaller than the median expected limit, as shown by the dot-dashed line in Fig.~\ref{fig:limitplot}. For these masses, the limit is constrained to a cross section such that the power of the alternate hypothesis is $\pi_{\rm crit}=0.16$~\cite{cowan2011power}.  This restricts the fluctuation to $1\sigma$ below the median expected limit. The introduction of the power constraint also introduces overcoverage, i.e., the coverage of the limit with the power constraint is greater than 90\%.
%The minimum of the limit curve is at $m_{\chi}=$~\SI{36}{GeV/c\squared} with a limit of $\sigma_{\mathrm{SI}}=$ \SI{9.2E-48}{cm\squared}.
The minimum of the limit curve is $\sigma_{\mathrm{SI}}=$ \SI{9.2E-48}{cm\squared} at $m_{\chi}=$~\SI{36}{GeV/c\squared}. The minimum of the unconstrained limit curve is \SI{6.2E-48}{cm\squared} at \SI{26}{GeV/c\squared}, and the minimum of the median expected limit is \SI{1.9E-47}{cm\squared} at \SI{43}{GeV/c\squared}.
The background model and data as a function of reconstructed energy are shown in Fig.~\ref{fig:energy}, and the data agree with the background-only model with a p-value of 0.96. LZ also reports the most sensitive limit on spin-dependent neutron scattering, detailed in the Appendix. A data release for this result is in the Supplemental Material~\footnote{See Supplemental Material at \url{\supplementalurl} for a data release for information shown in Figs.~\ref{fig:NRacc},~\ref{fig:s1s2wimp},~\ref{fig:limitplot},~\ref{fig:limitplotSDn}, and~\ref{fig:limitplotSDp}}.

The LZ experiment has achieved the highest sensitivity to spin-independent WIMP-nucleon scattering for masses greater than \SI{9}{GeV/c\squared} due to the successful operation of an integrated detector system containing the largest dual-phase xenon TPC to date. LZ is continuing operations at SURF and will undertake further detector and analysis optimization to search for a broad range of rare-event physics searches, including WIMPs, neutrinoless double-beta decay, solar neutrinos, and solar axions~\cite{akerib2019projected, akerib2020projected, akerib2021projected} over an estimated \SI{1000}{day} exposure.

The research supporting this work took place in part at SURF in Lead, South Dakota. Funding for this work is supported by the U.S. Department of Energy, Office of Science, Office of High Energy Physics under award no. DE-AC02-05CH11231, DE-SC0020216, DE-SC0012704, DE-SC0010010, DE-AC02-07CH11359, DE-SC0012161, DE-SC0015910, DE-SC0014223, DE-SC0010813, DE-SC0009999, DE-NA0003180, DE-SC0011702, DE-SC0010072, DE-SC0015708, DE-SC0006605, DE-SC0008475, DE-SC0019193, DE-FG02-10ER46709, UW PRJ82AJ, DE-SC0013542, DE-AC02-76SF00515, DE-SC0018982, DE-SC0019066, DE-SC0015535, DE-SC0019319, DE-AC52-07NA27344, \& DOE-SC0012447.	This research was also supported by U.S. National Science Foundation (NSF); the UKRI’s Science \& Technology Facilities Council under grant no. ST/M003744/1, ST/M003655/1, ST/M003639/1, ST/M003604/1, ST/M003779/1, ST/M003469/1, ST/M003981/1, ST/N000250/1, ST/N000269/1, ST/N000242/1, ST/N000331/1, ST/N000447/1, ST/N000277/1, ST/N000285/1, ST/S000801/1, ST/S000828/1, ST/S000739/1, ST/S000879/1, ST/S000933/1, ST/S000844/1, ST/S000747/1, ST/S000666/1, ST/R003181/1; Portuguese Foundation for Science and Technology (FCT) under grant no. PTDC/FIS-PAR/2831/2020; the Institute for Basic Science, Korea (budget no. IBS-R016-D1). We acknowledge additional support from the STFC Boulby Underground Laboratory in the U.K., the GridPP~\cite{faulkner2005gridpp,britton2009gridpp}  and IRIS collaborations, in particular at Imperial College London and additional support by the University College London (UCL) Cosmoparticle Initiative. We acknowledge additional support from the Center for the Fundamental Physics of the Universe, Brown University. K.T. L. acknowledges the support of Brasenose College and Oxford University. The LZ Collaboration acknowledges key contributions of Dr. Sidney Cahn, Yale University, in the production of calibration sources. We acknowledge Martin Hoferichter and Achim Schwenk for useful discussions. This research used resources of the National Energy Research Scientific Computing Center, a DOE Office of Science User Facility supported by the Office of Science of the U.S. Department of Energy under award no. DE-AC02-05CH11231. We gratefully acknowledge support from GitLab through its GitLab for Education Program. The University of Edinburgh is a charitable body, registered in Scotland, with the registration number SC005336. The assistance of SURF and its personnel in providing physical access and general logistical and technical support is acknowledged. We acknowledge the South Dakota Governor's office, the South Dakota Community Foundation, the South Dakota State University Foundation, and the University of South Dakota Foundation for use of xenon. We also acknowledge the University of Alabama for providing xenon.  We thank the journal reviewers for careful examination of our work, leading to some changes in our analysis and a more robust result. This work has been published as J. Aalbers et al. (LUX-ZEPLIN Collaboration)
\href{https://doi.org/10.1103/PhysRevLett.131.041002}{Phys. Rev. Lett. \textbf{131}, 041002 (2023)} under the terms of the Creative Commons Attribution 4.0 International license.

%For the purpose of open access, the authors have applied a Creative Commons Attribution (CC BY) licence to any Author Accepted Manuscript version arising from this submission.

\appendix
\section{Appendix - Spin Dependent Results}

WIMP-nucleon scattering can also have a spin-dependent interaction in which two limiting cases are considered: that WIMPs scatter only on protons or only on neutrons. Two isotopes of xenon have non-zero nuclear spin, \XeOneTwoNine\ (spin 1/2, \SI{26.4}{\percent} natural abundance) and \XeOneThreeOne\ (spin 3/2,  \SI{21.2}{\percent} natural abundance)~\cite{atomicweight}. As both have an unpaired neutron, the search is most sensitive to WIMP-neutron scattering. Sensitivity to a spin-dependent WIMP-proton interaction arises from mixing between proton and neutron spin states in isotopes with an unpaired neutron, albeit with increased uncertainty on the predicted signal rate~\cite{Engel:1991wq, klos2013structure, PhysRevLett.128.072502, PhysRevC.89.065501, PhysRevD.102.074018, Pirinen:2019gap}. 
Signal models for both the neutron and proton cases are constructed using the nuclear structure factors with uncertainties from Refs.~\cite{PhysRevD.102.074018,Pirinen:2019gap,PhysRevLett.128.072502}. This analysis quotes nominal limits which correspond to the mean structure functions from~\cite{PhysRevD.102.074018} and is chosen to facilitate a like-for-like comparison to previous limits from Xe-based experiments. An uncertainty is constructed for each $m_{\chi}$ by calculating the limit corresponding to the minimum and maximum interaction rate at each energy across the three models; this uncertainty also applies to the previous xenon results.
The details of data selection, background modeling, and statistical inference are identical to those reported in the main text.

\begin{figure}[htbp]
  \centering
  \includegraphics[ width=\linewidth]{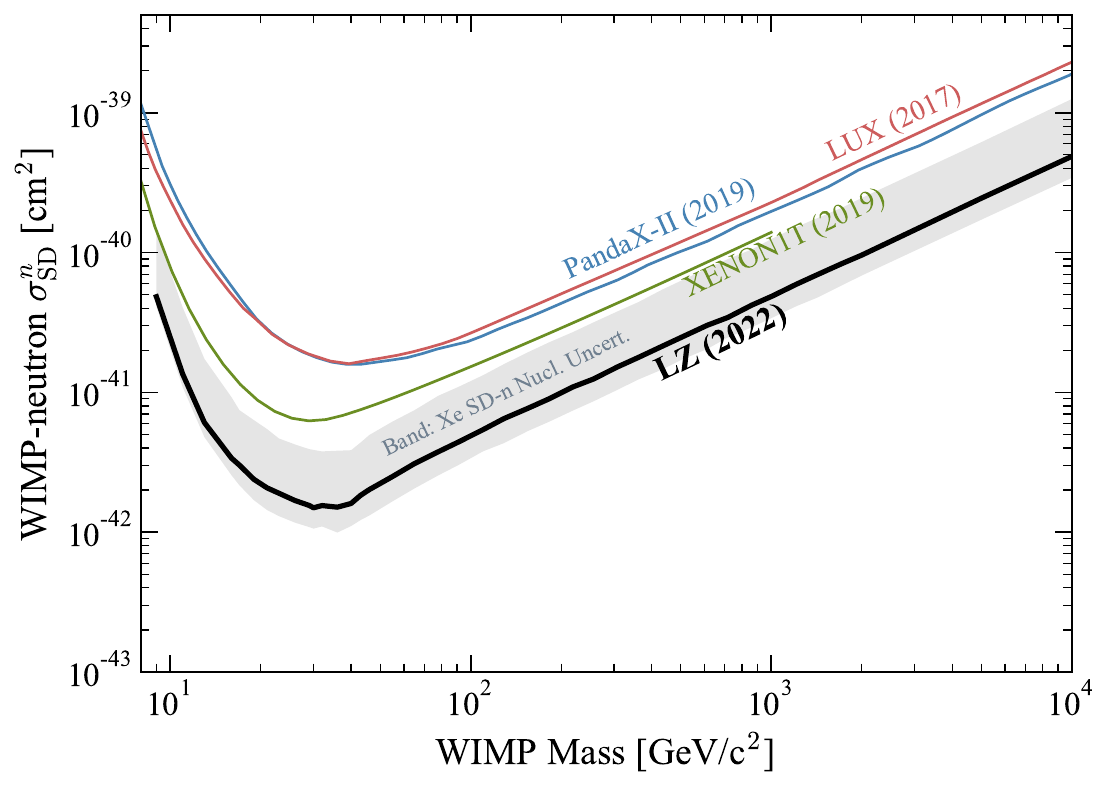}
  \caption{The \SI{90}{\percent} confidence limit (black line) and uncertainty bands (gray) coming from xenon nuclear correction factors for the spin-dependent WIMP-neutron cross section vs.~WIMP mass using the mean of the nuclear structure factors from~\cite{PhysRevD.102.074018} and range  across~\cite{PhysRevD.102.074018,Pirinen:2019gap,PhysRevLett.128.072502}. Also shown are the PandaX-II~\cite{xia2019pandax}, LUX~\cite{akerib2017limits}, and XENON1T~\cite{collaboration2019constraining} limits. A similar uncertainty band as shown on this result applies to the other Xe-based limits.}
  \label{fig:limitplotSDn}
\end{figure}

\begin{figure}[htbp]
  \centering
  \includegraphics[ width=\linewidth]{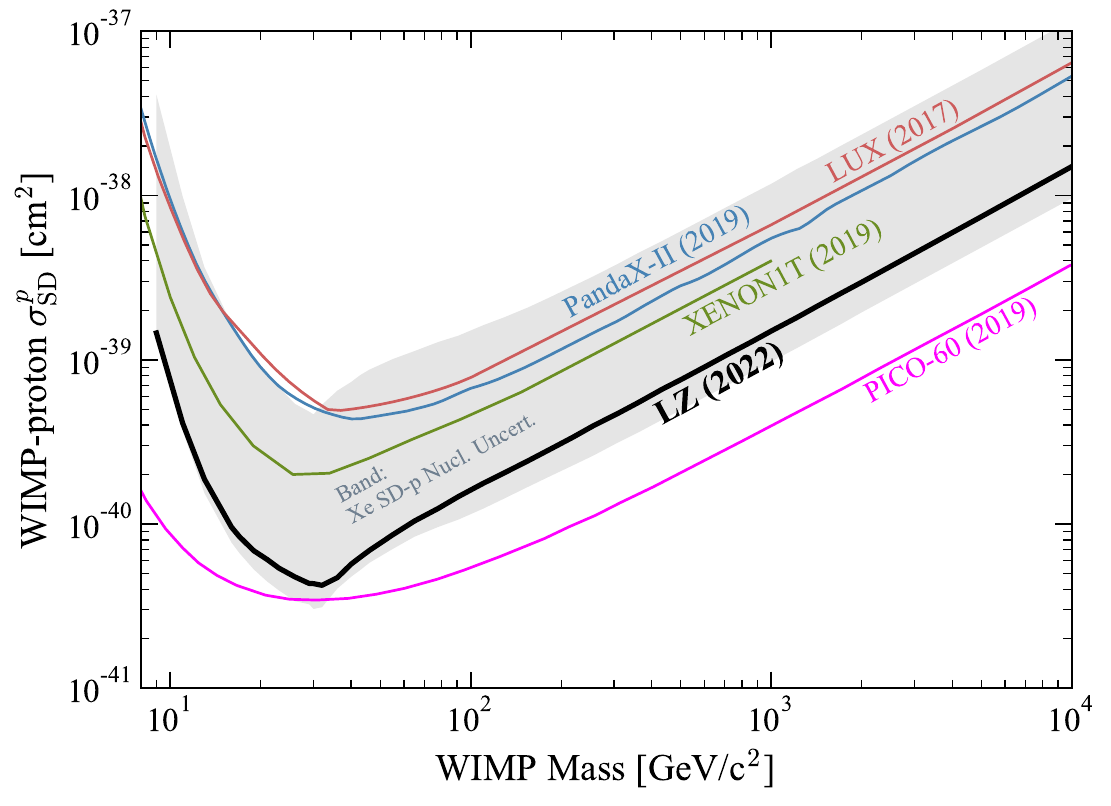}
  \caption{The \SI{90}{\percent} confidence limit (black line) and uncertainty bands (gray) coming from xenon nuclear correction factors for the spin-dependent WIMP-proton cross section vs.~WIMP mass using the mean of the nuclear structure factors from~\cite{PhysRevD.102.074018} and range  across~\cite{PhysRevD.102.074018,Pirinen:2019gap,PhysRevLett.128.072502}. Also shown are the PICO-60~\cite{amole2019dark}, PandaX-II~\cite{xia2019pandax}, LUX~\cite{akerib2017limits}, and XENON1T~\cite{collaboration2019constraining} limits. A similar uncertainty band as shown on this result applies to the other Xe-based limits. The PICO-60 result relies on WIMP-scattering on the spin of the unpaired proton of $^{19}$F with minimal uncertainty.}
  \label{fig:limitplotSDp}
\end{figure}

Above the smallest tested WIMP mass of \SI{9}{Gev/c\squared}, the best-fit number of WIMP events is zero for both neutron and proton cases, and the data are thus consistent with the background-only hypothesis. Figure~\ref{fig:limitplotSDn} shows the 90\% confidence level nominal upper limit (black line) and nuclear structure function uncertainty on the limit (grey band) on the WIMP-neutron spin-dependent cross section as a function of mass. The minimum of the limit curve is at $m_{\chi} =$~\SI{30}{GeV/c\squared} at a cross section of $\sigma_{\mathrm{SD}}^{n} =$ \SI{1.49E-42}{cm\squared};  a power constraint is applied between \SI{13}{GeV/c\squared} and \SI{36}{GeV/c\squared}.  Figure~\ref{fig:limitplotSDp} shows the 90\% confidence level nominal upper limit and uncertainty on the WIMP-proton spin-dependent cross section as a function of mass. The minimum of the limit curve is at $m_{\chi} =$~\SI{32}{GeV/c\squared} at a cross section of $\sigma_{\mathrm{SD}}^p =$ \SI{4.2E-41}{cm\squared}; a power constraint is applied between \SI{13}{GeV/c\squared} and \SI{32}{GeV/c\squared}. The minimum and maximum limits which form the nuclear structure factor uncertainty are also power-constrained over the relevant mass range for both the neutron and proton cases.

\bibliographystyle{apsrev4-2}
\bibliography{references}% Produces the bibliography via BibTeX.

%\end{document}

%%%%%%%%%% Merge with supplemental materials %%%%%%%%%%
\clearpage
\pagebreak

\widetext
\begin{center}
\textbf{\large Supplemental Materials}
\end{center}
%%%%%%%%%% Merge with supplemental materials %%%%%%%%%%
%%%%%%%%%% Prefix a "S" to all equations, figures, tables and reset the counter %%%%%%%%%%
\setcounter{equation}{0}
\setcounter{figure}{0}
\setcounter{table}{0}
\setcounter{page}{1}
\makeatletter
\renewcommand{\theequation}{S\arabic{equation}}
\renewcommand{\thefigure}{S\arabic{figure}}
\renewcommand{\thetable}{S\arabic{table}}
\renewcommand{\bibnumfmt}[1]{[S#1]}
\renewcommand{\citenumfont}[1]{S#1}
%%%%%%%%%% Prefix a "S" to all equations, figures, tables and reset the counter %%%%%%%%%%

\section{Detailed Event Rates}

\begin{table}[htbp]
    \caption{Number of events remaining after each stage of event selection criteria described in the main text.}
    \label{tab:events}
    \centering
    \begin{tabular}{cc}
    \\
    \hline
    \hline
    Selection description & Events after selection \\
    \hline
    All triggers & \num{1.1e8}\\
    Analysis time hold-offs & \num{6.0e7}\\
    Single scatter & \num{1.0e7}\\
    Region-of-interest & \num{1.8e5} \\
    Analysis cuts for accidentals & \num{3.1e4}\\
    Fiducial volume & \num{416}\\
    OD and Skin vetoes & \num{335}\\
    \hline
    \hline
    \end{tabular}

\end{table}

\section{Tuned Detector and Xenon Response Model Details}

The LZ detector and xenon response models are implemented in a \textsc{nest}-based application that includes effects such as curved electron drift paths from field non-uniformities, finite position reconstruction resolution in the transverse $(x,y)$ and longitudinal $z$ directions, and position-dependence in S1 and S2 areas. The key numerical parameters of the \textsc{nest} model are provided in Table~\ref{tab:NEST}. Additionally, a header file for \textsc{nest} 2.3.7 that will reproduce the ER and NR response models used in this analysis is available online at \url{\supplementalurl}. Note that the extraction field number is known to be an effective value due to multiple models for this effect in \textsc{nest}, and this parameter is tuned such that the extraction efficiency matches the LZ data.

In addition to the parameters below, the width of the predicted ER and NR bands had to be reduced to match LZ calibration data and, as mentioned in the main text, the \textsc{nest} recombination skewness model was turned off. There are detailed instructions for implementing these changes in the provided header file.

\begin{table}[htbp]
    \caption{\textsc{nest} tuning parameters. Parameters in the top half of the table are input parameters, while bottom half parameters result from \textsc{nest} calculations.}
    \label{tab:NEST}
    \centering
    \begin{tabular}{cc}
    \\
    \hline
    \hline
    Parameter & Value \\
    \hline \\[-2.2ex]
    $g_1^{\text{gas}}$ & \SI{0.0921}{phd/photon}\\
        $g_1$ & \SI{0.1136}{phd/photon}\\
    Effective gas extraction field & \SI{8.42}{kV/cm} \\
    \hline
    \hline
    Single electron & \SI{58.5}{phd} \\
    Extraction Efficiency & \SI{80.5}{\percent}\\
    $g_2$ & \SI{47.07}{phd/electron}\\
    \hline
    \hline
    \end{tabular}

\end{table}

\section{Limits in terms of number of WIMP events}

\begin{figure}[htbp]
  \centering
  \includegraphics[ width=0.5\linewidth]{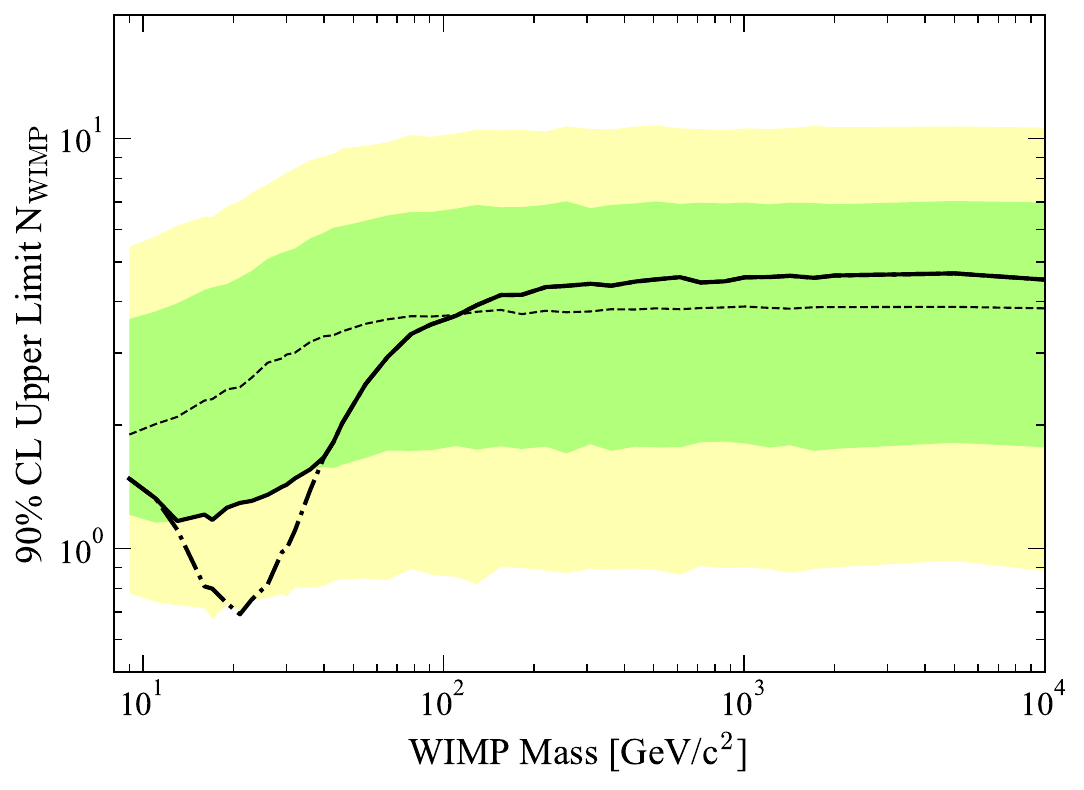}
  \caption{The \SI{90}{\percent} confidence limit (black line) for the number of WIMP events vs.~WIMP mass in the spin-independent case. The dot-dash line shows the limit before applying the power constraint described in the text. The dotted black line shows the median sensitivity. The green and yellow bands are the $1\sigma$ and $2\sigma$ sensitivity bands. }
  \label{fig:nwimp}
\end{figure}

The main text reports the limit in the standard form of WIMP cross section vs WIMP mass. In the supplemental material, we provide an alternative view of the result, reporting the limit as the number of WIMP events, for the spin-independent case, as shown in Fig.~\ref{fig:nwimp}.  As described in the main text, the limit is constrained by the power constraint method between \SI{13}{GeV/c\squared} and \SI{36}{GeV/c\squared}. Both the limit before the power constraint and the limit after the power constraint are shown, as well as the expected sensitivity. 

In unified interval analyses such as this, statistical fluctuations in the data create statistical fluctuations in the limit, including to values below the median sensitivity of the experiment; a further discussion of this effect is found in Section VI of~\cite{feldman1998unified}. We report the median expected limit, as well as the 1 and 2$\sigma$ bands around the median, as recommended, to mitigate the cognitive impact of these fluctuations. When comparing experiments or interpreting the compatibility of theory with this result, we recommend consideration of the median expected limit as well.

\section{Data Release}

Selected data from the following plots from this paper are available at \url{\supplementalurl}.

\begin{itemize}
    \item Figure~\ref{fig:NRacc}: points representing the total efficiency curve for this analysis (black line).
    \item Figure~\ref{fig:s1s2wimp}: points in S1-S2 space representing the data used in the WIMP search (black points).
    \item Figure~\ref{fig:limitplot}: WIMP mass points with measured $\sigma_{\mathrm{SI}}$ 90\% confidence limits and median and 1 and 2 sigma sensitivity bands.
    \item Figure~\ref{fig:limitplotSDn}: WIMP mass points with measured $\sigma_{\mathrm{SD}}^{n}$ 90\% confidence limits with uncertainty bands, and median and 1 and 2 sigma sensitivity bands. The sensitivity bands are not show in the plot for clarity.
    \item Figure~\ref{fig:limitplotSDp}: WIMP mass points with measured $\sigma_{\mathrm{SD}}^{p}$ 90\% confidence limits with uncertainty bands, and median and 1 and 2 sigma sensitivity bands. The sensitivity bands are not show in the plot for clarity.
\end{itemize}

\end{document}